\documentstyle[11pt,aasms4]{article} 

\slugcomment{Draft processed \today}

\lefthead{Reach and Rho}
\righthead{Excitation of a Molecular Cloud by SNR 3C~391} 

\begin{document}
 
\title{Excitation and Disruption of a Giant Molecular Cloud by the 
Supernova Remnant 3C~391}

\author{William T. Reach\altaffilmark{1}}
\affil{Infrared Processing and Analysis Center,
California Institute of Technology,
Pasadena, CA 91125}
\author{Jeonghee Rho\altaffilmark{2}}
\affil{
Service d'Astrophysique, CEA, DSM, DAPNIA, Centre d'Etudes de Saclay,
F-91191 Gif-sur-Yvette cedex, France}

\altaffiltext{1}{formerly:
Institut d'Astrophysique Spatiale, B\^atiment 121, Universit\'e
Paris XI, 91405 Orsay cedex, France}
\altaffiltext{2}{currently:
Physics Department, University of California, Santa Barbara, CA
}

\def\etal{ et al. }
\def\ISO{ {\it ISO} }
\def\arcsec{\hbox{$^{\prime\prime}$}}
\def\kms{ {km~s$^{-1}$}}
\def\simgt{\mathrel{\mathpalette\oversim>}}
\def\simlt{\mathrel{\mathpalette\oversim<}}
\def\cotwo{ $^{12}$CO($2\rightarrow 1$)}
\def\coone{ $^{12}$CO($1\rightarrow 0$)}
\def\cstwo{ CS($2\rightarrow 1$)}
\def\csthree{ CS($3\rightarrow 2$)}
\def\csfive{ CS($5\rightarrow 4$)}
\def\hcop{ HCO$^+$($1\rightarrow 0$)}

\begin{abstract}
Using the IRAM 30-m telescope,
we observed the supernova remnant 3C~391 (G31.9+0.0)
and its surroundings in the
\cotwo, \hcop, \cstwo, \csthree, and \csfive\ lines.
The ambient molecular gas at the distance (9 kpc) of the remnant
comprises a giant molecular
cloud whose edge is closely parallel to a ridge of bright
non-thermal radio continuum, which evidently delineates the blast-wave
into the cloud.
We found that in a small (0.6 pc) portion of the radio shell, the
molecular line profiles consist of a narrow (2 \kms) component,
plus a very wide ($> 20$ \kms) component.
Both spectral components peak within $20^{\prime\prime}$ of a previously-detected
OH 1720 MHz maser.
We name this source
3C~391:BML (broad molecular line); 
it provides a new laboratory, similar to IC 443 but on a larger
scale, to study shock interactions with dense molecular gas.
The wide spectral component is relatively brighter in the higher-excitation 
lines. We interpret the wide spectral component as
post-shock gas, either smoothly accelerated or partially dissociated and
reformed behind the shock. The narrow component is either the pre-shock
gas or cold gas reformed behind a fully dissociative shock.
Using the 3 observed CS lines, we measured the temperature,
CS column density, and H$_2$ volume density
in a dense clump in the parent molecular cloud
as well as the wide-line and narrow-line portions of the shocked clump.
The physical conditions of the narrow-line gas are
comparable to the highest-density clumps in the giant molecular cloud, while
the wide-line gas is {\it both} warmer and denser.
The mass of compressed gas in 3C~391:BML is high enough that its self-gravity 
is significant, and eventually it could form one or several stars.
\end{abstract}                            

\keywords{supernova remnants, ISM: individual (3C~391), ISM: structure,
radio lines: ISM}

\section{Introduction}

Supernovae are thought to be the source of kinetic energy of the
interstellar medium, keeping the gas in motion and returning material
from dense molecular clouds into the more diffuse interstellar medium
and the galactic halo.  When a massive star ends its life in a
supernova explosion, it often does so in the vicinity of the molecular
cloud in which it was born, as is evidenced by the close
correspondence of OB associations and giant H~II regions in spiral
arms (\cite{elmlad77}).  Despite the expected close association between Type II
supernovae and molecular clouds, very few cases of supernova-molecular
cloud (SN-MC) interaction are known or suspected.
The blast wave from a supernova within or near the edge of a cloud
will progress rapidly through the inter-clump medium and drive slower
shocks into dense clumps.  
Multiple reflections of high-energy charged
particles within the complicated magnetic field of an SN-MC
interaction are a possible source of cosmic rays, which will permeate
the entire region (\cite{chevalier77}; \cite{esposito96}).
The thermal radiation from the remnant interior
(mainly X-rays), cosmic rays and their secondary gamma rays, and
direct impact of the blast wave onto clumps should visibly perturb the
excitation, chemistry, and dynamics of the parent molecular cloud for
at least the $\sim 10^5$ year period during which the SN blast wave is
most powerful.

So far, the only well-known case of an SN-MC interaction is IC 443,
where molecular lines have been detected with FWHM $\sim$ 20 km s$^{-1}$ 
(much wider than the lines from nearby, un-shocked gas), and
from energy levels far above the ground state 
(\cite{white87}; van Dishoeck, Jansen, \& Phillips 1993; \cite{wang92}). 
Other remnants, including W~28, CTB 109, Kes 79, 
and W~51C have been suggested as
SN-MC interactions based on their proximity to 
molecular clouds, wide molecular lines, or both 
(\cite{woot77}; \cite{woot81}; \cite{tatematsu90}; \cite{green92};
\cite{koo97a}; \cite{koo97b}).  
In the case of W~28, W~44 and 3C~391, 
1720 MHz OH emission has been detected from many small spots,
with brightness temperatures
so high that they must be masers;
these masers are thought to be collisionally excited and
they strongly suggest the presence of SN-MC interactions
(Frail, Goss, \& Slysh 1994; \cite{frail96}).
3C~391 is one of the brightest radio supernova remnants, and high-resolution
radio images suggest a `break-out' morphology due to an
explosion near the edge of a molecular cloud (\cite{rm93}).
The X-ray emission from 3C~391 peaks in its interior and has a thermal spectrum
(\cite{rp96}), 
characteristic of a newly-defined class of supernova remnants,
called `mixed-morphology' remnants,
whose nature has been linked to interaction with a strongly inhomogeneous
pre-shock interstellar medium (\cite{rp98}). 
A recent map of the \coone\ emission in the vicinity of 3C~391 
revealed a giant molecular cloud that is precisely parallel to the bright
ridge of radio emission, confirming that its `break-out' radio morphology
is indeed due to the strong density contrast between the molecular cloud to
the northwest and the relatively empty regions elsewhere (Wilner, Reynolds, \& 
Moffett 1998). 
The work described in this paper is part of our recently-initiated campaign to
search for and characterize SN-MC interactions in the mixed-morphology
supernova remnants. Our first result was
the detection of bright [O~I] 63 $\mu$m and dust emission from 3C~391,
showing that the blast-wave into the molecular gas is radiative, and the SN-MC 
interaction is a significant energy loss for the remnant,
although it remains globally adiabatic (\cite{reach96}).
In this paper, we present new observations of
molecular emissions from 3C~391, designed to search for the effects of
the SN-MC interaction on the molecular cloud, using millimeter-wave observations
at high angular resolution and 
several transitions requiring a range of physical conditions for excitation.
Throughout this paper, we assume a distance to
3C~391 of 9~kpc, which is based on the comparison of H~I 21-cm emission 
and absorption line profiles (\cite{radakrish})
and the H$_2$CO absorption line at 96 \kms\
(\cite{downes}); our adopted distance is consistent with that adopted
by others for this remnant (\cite{rm93}).

\section{Observations and Results}  

\placetable{tab:telparams}

The observations described here were made with the IRAM
30-m telescope during 22--27 November 1996, in very good weather.
Three receivers were used to simultaneously observe three different spectral lines
with the 1.3, 2, and 3 mm receivers.
Our most-used receiver configuration contained the \cotwo,
\csthree, and \cstwo\ lines.
Part of the time, we switched the 1.3 mm receiver to the \csfive\ line,
and part of the time, we switched the 3mm receiver to the \hcop\ line.
The transitions, frequencies, telescope efficiencies, and beam-widths
(from \cite{wild}) are shown in Table~\ref{tab:telparams}.
Both the autocorrelator and filter-banks were used as back-ends, the former allowing
good velocity resolution and the latter used to cover a wide range of velocities.
Wide velocity coverage was particularly important so that we could search
for very broad lines and differentiate frequency-dependent gain variations 
of the receiver from the true spectral shape of our source.
Positions in the this paper will sometimes be expressed as offsets, 
relative to a central 
position (equinox 1950) of $18^{\rm h}46^{\rm m}50^{\rm s}$ $-1^\circ00^\prime00^{\prime\prime}$.
Each spectrum is the brightness relative to a reference position, which
was offset from the central position by $+255^{\prime\prime},-300^{\prime\prime}$
in r.a., dec, respectively. It was not possible to find a reference position close
to the remnant that was devoid of CO line emission, but using the
\coone\ map (\cite{wilner98}), we were able to find a position
with very low total emission and no emission at all at the velocities appropriate
for gas at the distance of 3C~391. 
There is a negative dip in the \cotwo\ spectra at 40 \kms\ due to contamination from 
the reference position.  All spectral lines
at velocities lower than 50 km s$^{-1}$ are not related to
the remnant and can be ignored. 

\placefigure{fig:comapamb}

\section{Distribution of Ambient Molecular Gas near 3C~391}
                                      
We mapped the entire remnant and surroundings on a
$20^{\prime\prime}$ grid, and then we mapped the region where we found wide
molecular lines on a $10^{\prime\prime}$ grid. 
The relationship between the nonthermal radio emission of the supernova remnant 
and the overall
distribution of molecular gas at the same distance is shown in Figure~\ref{fig:comapamb}.
The \cotwo\ line was integrated
over the velocity range 91--99 \kms, and the contours of radio continuum surface brightness
(from \cite{rm93}) are superposed. 
This LSR velocity range corresponds to gas at the distance of 3C~391 and 
participating in galactic rotation.
The CO emission is very bright to the northwest
of the remnant, with the edge of the bright region running precisely along the
radio shell of the supernova remnant.
The \coone\ emission, mapped over a larger area by Wilner \etal (1998),
shows that the molecular cloud boundary continues outside the boundaries of our map.
Looking on even larger scales, in the Massachusetts-Stony Brook CO survey(\cite{clemens86}), 
the CO emission spans an irregular region about $15^\prime$ minutes across.
Part of this region was identified as a giant molecular cloud 
with $(l,b,v)=(32.00^\circ,0.00^\circ,98\,{\rm km~s})$, $(\sigma_l,\sigma_b,\sigma_v)=(0.22,0.09,3.0)$, and its virial
mass was estimated to be $4\times 10^5 M_\odot$ (\cite{solomon87}). Including other
molecular emission that seems associated, the total mass of the complex is perhaps
twice as high. 
The \cotwo\ and \cstwo\ maps,
and the \coone\ map of a larger region (\cite{wilner98}), 
all show that the bright northeastern radio shell of 3C~391 is nearly tangent to the 
surface of this molecular cloud.
The radio, X-ray, and mm-wave CO-line morphologies suggest that
3C~391 is the result of an explosion near the edge of a giant molecular cloud. 
We will call this cloud the `parent cloud', because it is likely that the progenitor
was a massive star that formed in or near the cloud, but this term is merely suggestive
for the purposes of this paper.

\placefigure{fig:specrad}

\placefigure{fig:specac}

Over nearly the entire region shown in Figure~\ref{fig:comapamb}, 
the detected lines are all {\it narrow}, with FWHM $< 3.5$ \kms. 
Two example spectra, toward the peak of the nonthermal radio shell and
a bright position in the parent molecular cloud, are shown in
Figures~\ref{fig:specrad} and~\ref{fig:specac}, respectively.
The \cotwo\ line is detected almost everywhere, and the
\cstwo\ and \csthree\ lines are detected inside the boundary 
of the parent molecular cloud, which is to say, in the northwestern corner of
the map (Fig.~\ref{fig:specac}b).
The parent cloud is not uniform---it contains several clumps, which are
more pronounced in the CS maps. We identified a bright clump in the parent
molecular cloud, at (-150$^{\prime\prime}$,220$^{\prime\prime}$), 
for comparison with the physical properties of the
shocked gas. The ambient cloud has relatively weak \csthree\ emission and
undetectable ($< 0.1$~K) \csfive\ emission.

\placefigure{fig:manyspec}

\section{Molecular Spectra of a Shocked Clump} 
                   
The molecular emission lines change radically from those of the parent molecular
cloud in one small region, near one of the two OH masers found in 3C~391 (\cite{frail96})
at $18^{\rm h} 46^{\rm m} 47.69^{\rm s}$, -$01^\circ 01^{\prime} 00.6^{\prime \prime}$.
Every molecular transition we observed was significantly broadened within a small
patch of sky, about $1^\prime$ in diameter.
The spectra taken towards two positions,
only $45^{\prime\prime}$ apart, are shown in Figure~\ref{fig:manyspec}.
There are two components of the spectral lines, wide and narrow, which have very 
different relative brightnesses in different transitions.
The wide component of the spectral lines in Figure~\ref{fig:manyspec} has a full width at half
maximum (FWHM) of 20 \kms, while the narrow component has a FWHM of only 1.7 \kms.
The \cotwo\ line integral toward this clump is
400 K~\kms, which is an order of magnitude larger than the typical
line integral from the parent molecular cloud. 
The central velocity of the {\it narrow} component 
is the same as that of the OH 1720 MHz maser line (\cite{frail96}),
but the center of the wide component is somewhat offset (but by much less than the
line width).
Because the properties of the molecular emission in this region are so very different 
from those typical of the molecular gas in the region, and the position coincides
with a shock-excited OH maser and the [O~I] 63 $\mu$m emission peak,
we conclude that this is a region where the supernova
blast-wave has had a significant impact on the molecular gas. 
We will therefore refer
to this region as the `shocked clump,' or 3C~391:BML (broad molecular line), 
and we will attempt to measure its physical
properties in more detail in the discussion below.

\placefigure{fig:comosaic}

The shape of the spectral lines changes rather abruptly in the shocked clump,
and these line shape variations offer further clues to the nature of
the emitting region.
The \cotwo\ line profiles on a 
$10^{\prime\prime}\times 10^{\prime\prime}$ grid
are shown in Figure~\ref{fig:comosaic}.
Both the wide and narrow components of the spectral lines are bright
in this region, but their distributions are significantly different.
We separated the wide and narrow components from each spectrum, in
order to show their distributions separately.
For the narrow component, we subtracted a linear baseline fitted in a small
spectral window within 2 \kms\ of the narrow-line peak; this `baseline' includes
the wide component at the velocity of the narrow component peak.
For the wide component, we calculated
the total line integral over 85 to 130 \kms\ and subtracted the narrow component line integral.
A map of the wide and narrow components of the \cstwo\ lines
is shown in Figure~\ref{fig:CSmap}.
The narrow component peaks to the southeast and the wide
component to the northwest of the OH maser position, which is shown with an asterisk.
The \cotwo\ results are similar, but with a more extended narrow-line region.
The \csfive\ line profiles show {\it only} the wide component, and
the \csthree\ profiles are intermediate between the higher and lower CS transitions.

\placefigure{fig:CSmap}

Both the narrow and wide components are clearly related to the SN-MC
interaction because they both peak in the same region, and we are
evidently resolving regions of very distinct physical conditions. 
However, the exact relationship between the wide and narrow components 
is not clear.
The narrow component could be due to pre-shock gas, a hypothesis motivated by the fact
that the dense clump in the parent molecular cloud has the
same spectral lines in about the same proportions.
Alternatively, the narrow component could be due to molecules that have reformed
in gas that has cooled behind a fully dissociative shock. This second
hypothesis is motivated by the fact that the narrow component is spatially
near the wide component, there are infrared lines indicating a dissociative
shock, and the velocity of the narrow component closely matches that of the
shock-excited 1720 MHz OH maser. However, the present observations do not
clearly distinguish between these hypotheses. Indeed, the cold region far
behind a dissociative shock {\it should} be difficult to discern from pre-shock gas.
                           
\section{Overall distribution of shocked molecular gas }

The region shown in Figures~\ref{fig:comosaic} and~\ref{fig:CSmap} covers only a small part of the
whole supernova remnant. In particular, it does not include the location where 
one might expect the strongest impact of the shock on the parent molecular 
cloud---specifically,
the ridge of very bright radio emission that runs precisely along the
edge of the molecular cloud. The \cotwo\ and \cstwo\ spectra of the bright radio
ridge are shown in Figure~\ref{fig:specrad}. There is no trace 
of the wide emission line that characterized the shocked clump in
Figure~\ref{fig:manyspec}. The \cotwo\ line profile at the radio shell
splits into three components, with widths (FWHM) of 2 to 4 \kms.
If there is emission from a wide spectral component, then it must
be more than 40 times weaker at the radio shell than toward the shocked
clump. Further, there is no CS or HCO$^+$ emission detected toward
the radio shell; {\it neither} the wide nor narrow components are present.

\placetable{tab:spectab}

In order to determine whether the spectrum toward the radio shell is
`normal', and to illustrate the properties of spectral lines from the
parent molecular cloud, we show the spectrum of a bright clump from the
parent molecular cloud in Figure~\ref{fig:specac}. This clump was
identified as a peak in the \cstwo\ map, and its
CO lines are some of the widest in the parent molecular cloud.
The broadening of the CO lines is almost certainly due to saturation:
the CO lines are very optically thick.
The CO component at 96 \kms\ is the one that actually peaks at this
position, and is the only one in the CS spectrum. 
The line brightnesses and widths are summarized in Table~\ref{tab:spectab}.
Comparing the molecular spectra of the radio shell, the
parent molecular cloud, and other molecular clouds from
the galactic plane (\cite{clemens86}), we see no evidence that there
is any interaction between the molecular gas and the supernova remnant,
despite the strongly suggestive arrangement of the radio
shell exactly along the face of the molecular cloud.

\placefigure{fig:comapshock}
\placefigure{fig:comapshockb}

It is possible, however, that the interaction between the shock and
the molecular cloud leaves little or no trace in the observed molecular
lines, and the shocked clump 3C~391:BML is a rare exception. We looked
for more subtle indications of interaction by making maps in each of
the main velocity components of the molecular lines. Note that only
toward the shocked clump is there clear evidence for a very wide spectral
line. Everywhere else, the \cotwo\ and \cstwo\ lines (where detected)
consist of a sum of relatively narrow, Gaussian-shaped components.
There are 3 distinct velocity components of the \cotwo\ and \cstwo\ lines
between 90 and 120 km s$^{-1}$ that appear in at least part of the 3C~391
region. One of the spectral components (whose map is shown in
Figure~\ref{fig:comapamb}) is the bulk of the ambient molecular
cloud, which dominates to the northwest of the remnant.
Maps of the other two components are shown in Figures~\ref{fig:comapshock} 
and~\ref{fig:comapshockb}.  

Inspecting these maps, there is some evidence for a widespread 
interaction between 3C~391 and its parent cloud, roughly following
the entire hemisphere of the remnant facing the molecular cloud.  
There is a long, nearly straight
filament in the 109 \kms\ map (Fig.~\ref{fig:comapshock}) that runs 
roughly along the inside of the
northeastern part of the bright radio shell, all the way to the
southeastern extent of the bright radio emission. A spur from this
filament includes the position of the second OH maser (\cite{frail96}),
whose position is indicated on Fig.~\ref{fig:comapshock}.
The velocity and width of the CO, CS, and OH lines are in good agreement.  
Based on the unusual morphology, running along the edge of the non-thermal
radio shell, and the agreement in position and velocity with a
shock-excited OH maser, we suspect that the filament 
is related to the SN-MC interaction, even though the width of the
spectral lines in the filament is similar to that of the ambient gas.

Figure~\ref{fig:comapshockb} shows the distribution of gas with
velocities near 105 \kms. This is the velocity at which the narrow 
component of the clump 3C~391:BML peaks, and it contains part of the wide
component as well.
The map is dominated by a bright
peak at the position of 3C~391:BML, 
but it has secondary peaks in the parent molecular cloud as well as some
diffuse emission connecting the parent molecular cloud to 3C~391:BML.
The diffuse emission has a rough
correspondence with the southwestern part of radio shell, and it
contains---both in position and central velocity---the region with the
strongest SN-MC interaction; therefore, this emission may also be 
related to the SN-MC interaction.
The overall distribution of the molecular gas in Figs.~\ref{fig:comapshock} 
and~\ref{fig:comapshockb}, could be due to a hemisphere of displaced gas, 
with the open end pointing away from the parent molecular cloud. 
In this picture, 3C~391 is a `CO-shell' remnant---the first
of its kind. The `shell', however, is neither spatially complete nor kinematically
coherent, which is not surprising considering that the ambient molecular
cloud contains a great deal of structure.


\section{Excitation of the Molecular Gas}

\subsection{Three-line analysis of the CS excitation}

Using the three observed transitions of CS, we can constrain the physical
conditions in the shocked gas. 
We compared the line brightnesses to 
Large Velocity Gradient (LVG) models for a range of gas temperature, $T$, 
CS column density, $N({\rm CS})$, and gas density, $n({\rm H}_2)$. 
We can reject the possibility that the observed lines are in
Local Thermal Equilibrium (LTE), because the observed CS line ratios imply
kinetic temperatures even lower than the observed CO brightness temperatures; that is,
the excitation of CS must be sub-thermal. This is not surprising, because
the critical density for collisional de-excitation of the \csfive\ line
is very large: $n_c \simeq 8\times 10^6$~cm$^{-3}$.
So far, we have only considered excitation by collisions with H$_2$ molecules,
which is probably sufficient for the ambient cloud. For the shocked gas,
some excitation by collisions with atomic H or even electrons is possible if 
the H$_2$ formation is incomplete or the ionizing photon field is strong; 
in this case, our $n({\rm H}_2)$ refers to a composite of the
H$_2$, H, and electron densities responsible for the CS excitation.
It is quite likely that the shocked gas is partially dissociated, so that
H is the dominant collision partner there, as was found in W~51C (\cite{koo97b}).
Based on previous work in interpreting similar CS observations of diffuse
clouds (\cite{reach95}), dipole selection rules mean electrons should be of 
relatively little importance
except for $1\rightarrow 0$ transitions, such as the bright \hcop\ lines;
the positive charge of HCO$^+$ also means electrons are likely to
be important in its excitation. 

\placefigure{fig:csexcite}

Model parameters consistent with the observations for two positions are
illustrated in Figure~\ref{fig:csexcite}. 
The first position is the strongly-shocked clump---wide spectral component only---for
which the spectra were shown in Figure~\ref{fig:manyspec}.
The second position is a dense clump in the parent molecular cloud,
(well {\it outside} the remnant), for which the spectra were shown in 
Figure~\ref{fig:specac}.
The temperature cannot be determined separately from the volume density for
the ambient gas, because only 2 lines were detected.
We assume a temperature of 20~K for the clump in the parent molecular cloud.
This temperature is consistent with the lower limit set by the observed brightness 
temperature of the \cotwo\ line, and it is in between the 
10--15~K inferred from 
NH$_3$ and CS observations of dark clouds (\cite{benson89}; Snell, Langer, \& 
Frerking 1982)
and higher temperatures found in star-forming cores from H$_2$CO
observations (\cite{mangum97}).
For reference, we note that in LTE the observed CS line ratio yields an
excitation temperature of only 3.7~K, which is clearly lower than the
gas temperature.
The contours of constant $3\rightarrow 2$/$2\rightarrow 1$ ratio and
$2\rightarrow 1$ brightness for the ambient clump are shown in Figure~\ref{fig:csexcite}.
These constraints intersect at
a volume density of $6\times 10^4$ cm$^{-3}$ and a CS column density of
$9\times 10^{12}$ cm$^{-2}$. Assuming we are observing a uniform clump with
depth equal to transverse size (comparable to the beam size), the abundance of 
CS relative to H$_2$ is $4\times 10^{-11}$.
We take these values to characterize the densest clumps in the parent cloud. 
There are of course wide density variations in the molecular cloud.
The \csthree\ emitting regions have densities $> 10^4$ cm$^{-3}$; they
are detected over most of the ambient cloud but probably do not fill its volume.


\subsection{Physical parameters of the shocked gas}

For the shocked gas, we found that low-temperature models are
not able to produce the \csfive\ emission in the observed amounts, even
at very high densities. A lower limit to the gas temperature is the
observed brightness temperature of the \cotwo\ line: $T_k>18$~K.
In the high-density limit (LTE), the observed line ratios yield $T_k=$7--13~K, depending 
on which pair of lines is used; this LTE solution is inconsistent with the
observed brightness temperature.
There is a family of LVG solutions that can approximately match the observed
line brightnesses. First, there is
a lower limit for the temperature of the shocked gas given by the lowest
temperature that can produce the observed ratio of $5\rightarrow 4$/$2\rightarrow 1$
brightness: $T_k > 40$~K.
Within the range 40--100~K, the models that match the observed $5\rightarrow 4$/$2\rightarrow 1$
ratio require densities in the range of 1--0.3$\times 10^{6}$ cm$^{-3}$.
The highest temperature that we considered in the LVG analysis, 100 K, gives
a rough match to all three CS lines. Higher temperatures will probably also work,
but our numerical model becomes inaccurate.
None of the models match all three lines, although agreement improves as the 
temperature increases.
The excitation of the shocked gas is also constrained by the relative \cotwo\ line brightness,
compared to \coone\ from Wilner \etal (1998). There is no wide component detected in the
\coone\ line, but we can limit it to $T_b(1\rightarrow 0)< 1.5$~K within the beam
of the 12-meter telescope. The relative filling factor of the wide component
in the 12-meter beam $\sim 0.3$ based on our map, so
we can limit the ratio of $1\rightarrow 0/2\rightarrow 1 < .3$. The CO lines,
unlike CS, are likely to be thermalized because of the lower dipole moment
of the former molecule; using LTE, the CO line ratio implies $T_k > 50$.
We conclude that the shocked clump is both significantly more {\it hot} and {\it dense}
than the densest clumps in the parent cloud,
while the narrow-line emitting region of the shocked clump is similar to a dense
clump in the parent molecular cloud. Therefore, there
 is an abrupt change in the physical properties
of the molecular-line emitting gas within 3C~391:BML.

The column density of CS is well determined in the shocked gas,
because we directly observed the column density of CS molecules in levels
$J=2$, 3, and 5, which include the most-populated levels.
For the shocked gas, $N({\rm CS})=3\times 10^{13}$~cm$^{-2}$.
If the wide-line emitting region is uniform and spherical, with depth equal
to the transverse size (1.1 pc) then
the column density of the collision partner of CS, presumably
H$_2$ but including H~I, is
$N\simeq$3--1$\times 10^{24}$~cm$^{-2}$, and the
CS abundance in the wide-line emitting region is
1--3$\times 10^{-11}$ for temperatures 40--100~K.
This CS abundance is about an order of magnitude {\it lower} than in 
the dense clump in the parent molecular cloud,
suggesting that the CS was dissociated and has not completely reformed 
behind the shock.
However, it is possible that the emitting region does not fill the beam, 
even though the emitting region is clearly resolved.
For a fast, dissociative shock, the thickness from the 
shock front to where the gas has cooled to 100 K is only predicted to be
of order 0.03 pc (\cite{hm89}), compared to the 0.6 pc extent of the region where
we observed the wide emission lines. 
The CS abundance in the 
shocked clump could therefore be as high as that of the dense clump
in the parent molecular cloud, and
the CS molecules could be nearly completely reformed behind the 
shock---or never dissociated at all.
If this is true, then the
wide-line emitting region would comprise a network of relatively thin
shock fronts that could be revealed in higher-resolution images.

The chemistry of the shocked gas is apparently not significantly different
from that of the parent molecular cloud, at least for the molecules
that we have observed. The CS abundances of the shocked clump and the
ambient, dense clump in the parent molecular cloud are approximately equal
(within a factor of 4 even in the most extreme case considered).
Compared to those measured for other molecular clouds (\cite{irvine}),
the measured abundances of CS, CO, and HCO$^+$ relative to H$_2$
(with H$_2$ inferred from the CS excitation) are low more than an
order of magnitude.
But the {\it relative} abundances between the molecules we actually observed are
very similar to that of dense, quiescent molecular gas. 
The relative abundances are nearly independent of the
kinetic temperature or density. We find
CS/CO$\simeq (1\pm 0.3)\times 10^{-4}$, and HCO$^+$/CS$\sim 0.5$.
These abundances are very similar both to those found for the
shocked gas in IC~443 (Ziurys, Snell, \& Dickman 1989; 
\cite{vandis93}) and W~51C (\cite{koo97b}), 
and to those of quiescent gas in
TMC~1 and the Orion ridge (\cite{irvine}). The fact that our abundances
are all low relative to H$_2$ (as inferred from the excitation of CS)
could be due to a combination of (1) the emitting region is structured
on smaller scales, (2) we have missed some of the molecules because we
only observed a limited number of transitions, and (3) there is a range
of physical conditions on the line of sight so that different lines
come from different regions. 
Assuming that atomic H is the dominant 
collision partner, rather than H$_2$, only helps by about a factor of about 2,
because 2 H atoms are unlikely to be much more effective than one H$_2$ 
molecule. 
The discrepancy between excitation and abundance is not unique to shocked gas; this
is a more general problem in understanding the excitation of molecular gas
(see, {\it e.g.}, \cite{falgarone98}). If ambient molecular gas is already
highly structured on very small scales, then it would not be surprising
for the post-shock gas to be also highly structured. 	

\subsection{Dynamics of the shocked gas}

The velocity of the shocks that compressed and heated 3C~391:BML can
be estimated in several ways. 
The width of the wide spectral component probably gives a lower limit to 
the shock velocity, 
$v_s > 20$ \kms,
because it is a one-dimensional projection onto
the line of sight. Furthermore the gas might not be detected along the
entire range of velocities between the pre-shock and post-shock gas,
as it may be dissociated part of the time.
Another estimate is obtained by comparison with 
the CO emission predicted for magnetohydrodynamic shocks by
Draine \& Roberge (1984):
the \cotwo\ line brightness we observed from the wide-line clump
($\sim$5$\times$10$^{-6}$erg s$^{-1}$cm$^{-2}$sr$^{-1}$) is
consistent with models for a pre-shock density of $10^4$--$10^5$ cm$^{-3}$ and
a range of shock velocities $v_s\simeq$ 10--50 \kms. 
This is consistent with the observed linewidth, if we are
observing shock fronts with a range of orientations relative to our line of sight.
The location of the 3C~391:BML near the edge of the remnant suggests that
the shock would be perpendicular to the line of sight, favoring higher
values of $V_s$.
This range of shock velocities is near the transition between dissociative (J) and
non-dissociative (C) shocks. 
In W~51C, Koo \& Moon (1997b) concluded that the shock was dissociative,
and the CO molecules have re-formed while the H remains largely atomic.
In contrast, we suspect that the wide emission lines in 3C~391:BML are from
a shock that is only partially dissociative.
This is consistent with the fact that the pre-shock density is substantially higher for
3C~391:BML than was estimated for W~51C.

A simple estimate of the ram pressure of the 3C~391 blast wave 
(energy per unit volume, assuming adiabatic expansion)
suggests that it drive shocks with $n_0 v_s^2\sim 3\times 10^5 E_{51}$,
where $n_0$ is the pre-shock density in cm$^{-3}$, $v_s$ is the shock velocity
in \kms, and $E_{51}$ is the SN energy in $10^{51}$ erg.
The combination of 
shock velocity and pre-shock density inferred from the millimeter
observations is an order of 
magnitude higher than our simple estimate for the ram pressure
of the blast wave.  
The same conclusion was reached from interpretation of the bright [O~I] 63 $\mu$m
lines from this remnant and W~44 (\cite{reach96}). 
A surprisingly high pressure ($\sim 25$ times) was also inferred by
Moorhouse et al. (1991) from an H$_2$ observation of IC~443.
We can consider a few possible explanations for why the spectral diagnostics yield
such high pressures.
The first possibility is that the pressure is higher due to reflected shocks. 
When a blast wave hits a high density structure,
the shock splits into reflected and transmitted shocks.
For a density ratio $\sim 10^3$ between the cloud and the inter-cloud medium, the pressure
is enhanced a factor of several in plane-parallel, uniform-density models
(Borkowski, Blondin, \& Sarazin 1992; Hester, Raymond, \& Blair 1994).
While a reflected shock does not appear to be fully sufficient,  
a model accounting for similar physical processes and the specific conditions for 3C~391 
(a globally adiabatic remnant impacting a giant molecular cloud) might yield higher
pressures.

Another scenario is gravitational binding of the clumps, after compression
by the strong shock. Assuming the dominant collision partner of CS is H$_2$,
and assuming the emitting region is a uniform sphere,
the mass of 3C~391:BML is of order $10^4 M_\odot$.
With this mass, the
gravitational pressure $\rho v_{esc}^2$, estimated using the escape velocity
at the edge (0.6 pc radius) 
is comparable to pressure inferred
from the molecular and [O~I] line brightnesses. 
The gravitational potential of the shocked clump is large enough that
only very significant turbulent motions could prevent collapse.
Thermal motions for $T\sim 200$~K and magnetic pressure for a field
strength $B_{los}\sim 200$~$\mu$G (as measured for a similar shocked
clump by Claussen et al. 1997) are negligible compared to self-gravity.
Random motions with a dispersion of $\sim 25$ km~s$^{-1}$ could balance
self-gravity. The observed line widths are comparable to this, but
it is not clear what fraction of the observed motions are bulk flow
and what fraction is random, or turbulent.
The pre-shock gas has random motions only of order 3 km~s$^{-1}$,
and the shock front will impart some fraction of the shock
velocity into turbulent motions. The shock-induced turbulence is
expected to decay rapidly (\cite{maclow98}), leaving the post-shock gas
vulnerable to gravitational collapse.
Our observations have too low spatial resolution
to determine whether the emitting region is highly 
fragmented---in which
case our uniform-density mass estimate could be too high, and
the individual fragments might be gravitationally stable or unbound.
However, taken at face value, a uniform emitting region would have a very
significant self-gravity.

We could be observing in 3C~391:BML the
triggering of star formation. Star formation could not have occurred yet, since
the free-fall time is longer than the remnant age.
We estimate that the shock hit the dense clump less than 1500 yr ago:
the clump is within $d<0.7$~pc of the nearest edge of the radio continuum shell, 
and $d = R - R_o = R_o [(t/t_o)^{2/5} - 1]$,
where $R$ and $t$ are the radius and age now, and $R_o$ and $t_o$ are
the radius and age of the remnant when it hit the clump.
A similar scenario was first considered by Elmegreen \& Lada (1977),
where shock waves trigger star formation by compressing adjacent molecular clouds.
In the case of 3C~391, the present configuration appears consistent with a single
explosion near the edge of a molecular cloud. The bright radio
continuum shell and the CO shell would correspond to the the layer
considered by Elmegreen \& Lada. This entire layer may not become 
gravitationally unstable from the action of a single remnant.
We found that most of the radio shell, including the intense radio ridge,
does not contain dense molecular gas. Where the shock front strikes a
particular dense clump, which is our intepretation of
3C~391:BML, that clump could collapse separately to form one or several stars.

\def\extra{
Finally, we briefly examine whether the CO lines from the shocked clump 
are masers. 
Based on Draine \& Roberge's (1984) analysis of CO excitation, it appears
that the level populations of CO can be inverted in shocks with the ram
pressures we are considering. This raises the possibility
that part of the molecular emission we are observing is conceivably
a maser. The maser amplification is predicted to be relatively small,
because the path length through the region with inverted levels is small,
but brightness temperatures of order 100 K are predicted. 
In fact we already observe a main-beam brightness temperature up to 50 K
in \cotwo\ with
a $10^{\prime\prime}$ beam; if the emission is significantly structured
on smaller angular scales, then
the brightness peaks could exceed the gas temperature.
However, the fact that the \csthree\ and \cstwo\ and
\hcop\ lines in Figure~\ref{fig:manyspec} all have similar shape and comparable
brightness, suggests that they, at least, are not masing.
Moreover, the fact the narrow-line region appears to be resolved, unless 
the emission is from many spots smaller much smaller than the beam.
We therefore think it is unlikely that the bright millimeter lines we observed
are masers.
Instead, we think it arises from compressed clumps.    
}

\section{Implications for molecular shocks in other supernova remnants}

Given the high brightness temperature and line-width of the emission from the
strongly-shocked clumps in 3C~391, IC~443, and W~51C,
why hasn't shocked gas been observed in other supernova remnants? 
First, the strongly-shocked gas occupies a very small region ($< 0.6$ pc).
Therefore, it will be either beam-diluted or easily missed
in low-resolution or incompletely-sampled maps.
Second, detection of the broad line requires a wide bandwidth and flat baseline,
both of which were possible in good observing conditions and using modern correlators.
A third reason that other shocked molecular clouds
like those in 3C~391 and IC443 have not been detected is that they are much less
prominent in lower-excitation lines like \coone. The shocked gas is comparable to
pre-shock gas in \cotwo\ and \cstwo\ emission and dominates for higher
excitation transitions.
Fourth, there is often confusion in the line profiles due to absorption by molecular
gas both from the parent molecular clouds and clouds along the line of sight.
In the shown spectra in Figure~\ref{fig:manyspec}, there are dips that could be
self-absorptions in the \hcop\ and \cotwo\ lines.
However, we were able to obtain a $^{13}$CO($1\rightarrow 0$) spectrum at this position
that showed that the cold gas peaks at the same velocity of the narrow component---not at
the velocity of the dips in the spectra.
The {\it lack} of self-absorption is one of the main reasons why
we can clearly see the shocked gas from 3C~391. 
For comparison with W~44, the complicated line profiles of the shocked gas left some doubt 
as to whether the emission is broad or just multiple components (\cite{denoyer83});
however, the dips in the \coone\ profile are correlated to peaks in the 
$^{13}$CO profile (\cite{seta96}), showing that they are self absorption.
We found that this relationship occurs in detail over several areas of
the supernova remnants W~44 and W~28, and will report these results in a later paper.
The more extended, narrow-line emission that we suggested is a `CO shell' for 3C~391
is not clearly
distinct from ambient molecular gas, and we have relied on morphological
agreement with the radio shell as well as position and velocity agreement with
the shock-excited OH masers.
Wide- and narrow-line shocked regions, and narrow-line ambient regions
are expected to coexist, until such a time
as the shock has completely passed through the all the dense clumps, leading to 
complicated line profiles. 
These factors are likely to be true for the other remnants interacting with molecular 
clouds, and we expect more detections of similar objects in the future.

\section{Conclusions}

We observed the entire supernova remnant 3C~391 in millimeter lines of CS, CO,
and HCO$^+$. The lower-excitation lines reveal a giant molecular cloud to the
northwest of the remnant, explaining the `break-out' morphology of the radio emission.
The interactions between the blast wave and a very dense molecular clump
was found within a small ($50^{\prime\prime}$) region that we call 3C~391:BML. 
A wide component (FWHM 25 \kms) and a narrow component (FWHM 2 \kms)
both peak at 3C~391:BML, which is coincident with an OH 1720 MHz maser.
The excitation of the wide molecular lines requires both high gas temperature
($> 100$ K) and density ($\sim 3\times 10^5$ cm$^{-3}$). The narrow-line region
require somewhat lower density and are consistent with much lower ($\sim 20$ K) temperatures.
We identified a clump in the parent molecular cloud with properties similar to
the narrow-line region. Therefore, the 3C~391:BML clump was similar to the highest-density
clumps in the parent molecular cloud, and it is currently being shocked.
The brightness of the wide \cotwo\ line from 3C~391:BML
is consistent with C-type molecular shocks with
$10^4 < n_0 <  10^5$ cm$^{-3}$ and $10 < v_s < 50$ \kms, or J-type shocks
with $n_0\sim 10^3$ and $v_s \sim 100$ \kms.
The pressure in the shocked clump is much higher than the estimated ram pressure of
the remnant, possibly because of its self-gravity. If so, this clump is a likely
site of triggered star formation.
A widespread interaction of 3C~391 with molecular gas is evidenced by the
distribution of CO and CS lines with central velocities offset from that of the parent
cloud by 10 to 15 \kms; these velocities correspond to those of the two OH masers.
The interaction comprises nearly an entire hemisphere of the remnant, making 3C~391
a `CO shell' remnant.  

{\bf Acknowledgment}
We thank Hans Ungerechts at IRAM for helping us get started on the IRAM telescope
and David Wilner and Steve Reynolds for sharing early results of their observations.
WTR thanks the Commissariat d'Energie Atomique, in Saclay, France for hospitality 
and computing power during part of the data analysis. We thank Bon-Chul Koo for
his comments and support. The research described in this paper was carried out
in part by the California Institute of Technology, under a contract with the
National Aeronautics and Space Administration.

\clearpage
 
\begin{deluxetable}{lllllll}
\footnotesize
\tablecaption{Observed spectral lines and Telescope parameters\label{tab:telparams}}
\tablewidth{0pt}
\tablehead{
\colhead{transition} & \colhead{frequency} & \colhead{$T_{sys}$\tablenotemark{a}} 
  & \colhead{$\eta_{mb}$} &  \colhead{beam} & 
  \colhead{$\delta v$\tablenotemark{b}}  &    \colhead{$\Delta v$\tablenotemark{c}} \\
                     & \colhead{(GHz)}     & \colhead{(K)} 
  &                           & \colhead{($^{\prime\prime}$)} & 
   \colhead{(\kms)}  & \colhead{(\kms)}  }
\startdata
HCO$^+$($1\rightarrow 0$) &  89.1885 & 310 & 0.82 & 27 & 0.22 & 430\nl
CS($2\rightarrow 1$)      &  98.9798 & 250 & 0.76 & 24 & 0.24 & 430\nl
CS($3\rightarrow 2$)      & 146.9690 & 240 & 0.58 & 16 & 0.65 & 290\nl
CS($5\rightarrow 4$)      & 244.9356 & 490 & 0.43 & 10 & 1.2  & 310\nl
CO($2\rightarrow 1$)      & 230.5380 & 720 & 0.45 & 10 & 1.3  & 330\nl
\enddata
\tablenotetext{a}{typical system temperature for observations presented in this paper}
\tablenotetext{b}{velocity resolution}
\tablenotetext{c}{velocity coverage}
 \end{deluxetable}

\clearpage

\begin{deluxetable}{llll}
\tablecolumns{4}
\footnotesize
\tablecaption{Measured properties of spectral lines\tablenotemark{a}\label{tab:spectab}}
\tablewidth{0pt}
\tablehead{
\colhead{transition} & \colhead{$T_{mb}$} 
  & \colhead{$V_{LSR}$} &  \colhead{$\Delta V$ (FWHM)} \\
                     & \colhead{(K)}  
  & \colhead{(\kms)}  & \colhead{(\kms)}  }
\startdata

\cutinhead{wide-line position in shocked clump $(-40^{\prime\prime},-50^{\prime\prime})$\tablenotemark{b}}
HCO$^+$($1\rightarrow 0$) & 0.91 & 111.4 & 25.5 \nl   
CS($2\rightarrow 1$)      & 0.32 & 108.9 & 19.0 \nl   
			  & 0.24 & 104.2 &  1.4 \nl   
CS($3\rightarrow 2$)	  & 0.37 & 108.9 & 20.0 \nl   
CS($5\rightarrow 4$)	  & 0.35 & 108.5 & 15.6 \nl
CO($2\rightarrow 1$)	  & 17.6 & 111.1 & 22.5 \nl
			  & 15.1 & 104.2 &  2.6 \nl

\cutinhead{narrow-line position in shocked clump $(-10^{\prime\prime},-85^{\prime\prime})$\tablenotemark{b} }
HCO$^+$($1\rightarrow 0$) &  1.0 & 105.3 &  1.7 \nl   
                          & 0.38 & 102.6 & 14.4 \nl   
CS($2\rightarrow 1$)      & 0.90 & 105.5 &  1.0 \nl   
                          & 0.23: & 103.7 &  6.0 \nl  
CS($3\rightarrow 2$)	  & 0.55 & 105.4 &  1.2 \nl   
                          & 0.23 & 103.7 &  6.0 \nl   
CS($5\rightarrow 4$)	  & $<0.13$ & & \nl           
CO($2\rightarrow 1$)	  & 11.6 & 105.5 & 1.2 \nl
                          &  9.7 & 104.8 & 3.7 \nl

\cutinhead{radio ridge $(-130^{\prime\prime},+90^{\prime\prime})$\tablenotemark{c} }
HCO$^+$($1\rightarrow 0$) & 0.22: &  96.4 &  2.4 \nl
CS($2\rightarrow 1$)      &  $<0.03$ & & \nl          
CS($3\rightarrow 2$)      &  $<0.04$ & & \nl          
CS($5\rightarrow 4$)      &  $<0.13$ & & \nl          
CO($2\rightarrow 1$)	  &  8.9 &  96.5 &  4.0 \nl
                          &  3.3 & 102.0 &  2.1 \nl
                          &  3.6 & 107.0 &  3.8 \nl

\cutinhead{clump in parent cloud $(-130^{\prime\prime},+240^{\prime\prime})$\tablenotemark{c} }
HCO$^+$($1\rightarrow 0$) &  0.32 &  96.1 & 7.3: \nl 
CS($2\rightarrow 1$)      &  0.55 &  96.7 & 3.8 \nl
CS($3\rightarrow 2$)	  &  0.27 &  96.5 & 3.6 \nl
CS($5\rightarrow 4$)	  & $<0.13$ & & \nl          
CO($2\rightarrow 1$)	  & 12.0 &  96.0 & 5.4 \nl
                          &  8.4 & 107.9 & 4.0 \nl
\enddata
\tablenotetext{a}{only components between 90 and 140 \kms\ are listed}
\tablenotetext{b}{spectra averaged within $11^{\prime\prime}$ radius}
\tablenotetext{c}{spectra averaged within $21^{\prime\prime}$ radius}
 \end{deluxetable}

\clearpage

\figcaption[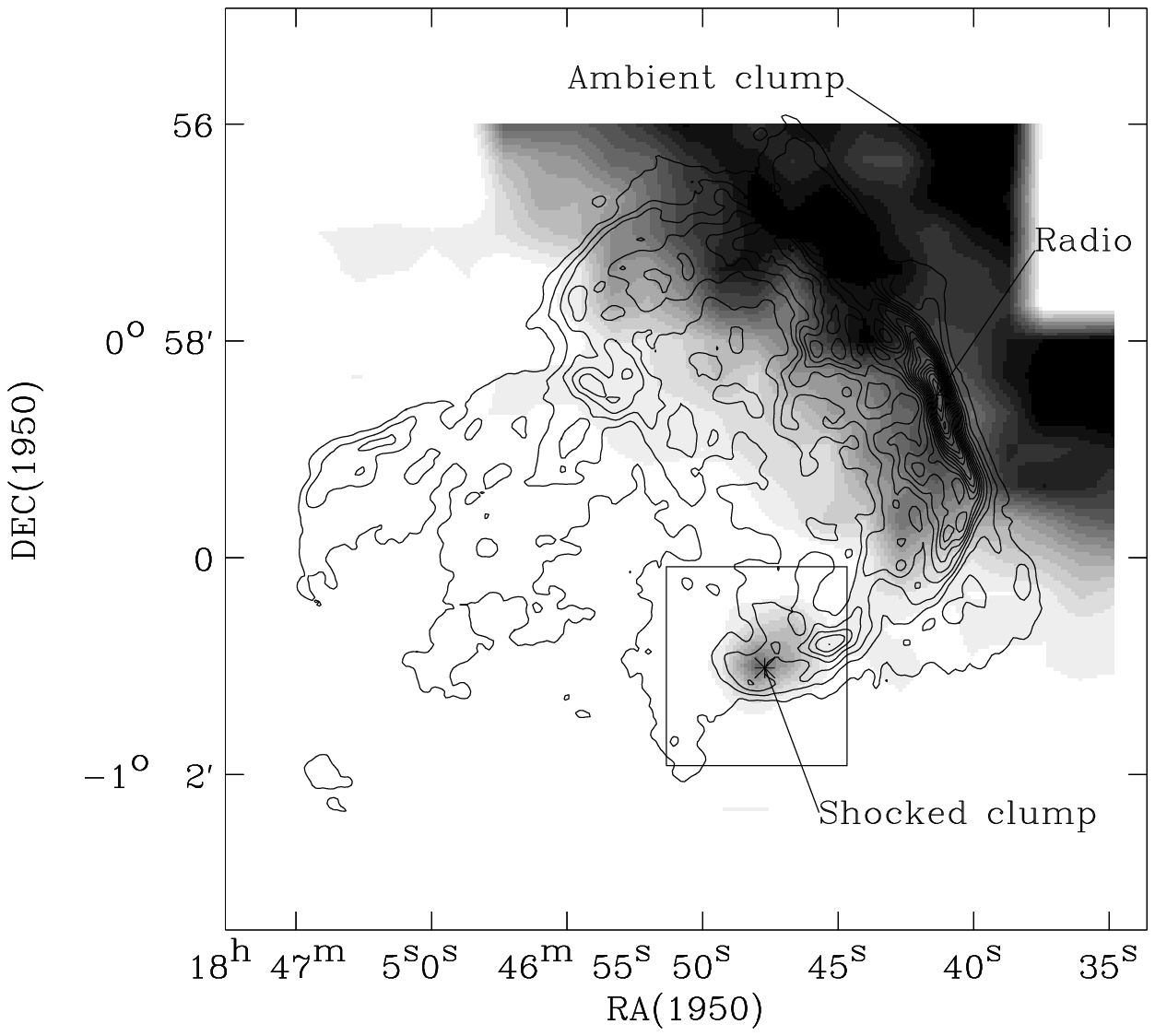]{An greyscale overlay of the \cotwo\ brightness 
(integrated over the velocity
range 91--99 \kms) on the radio continuum contours. 
The CO observations are sampled every
20$^{\prime\prime}$. The region of bright CO emission to the northwest is  
the likely parent cloud of the 3C~391 supernova progenitor.
The positions of the `radio shell,' `ambient clump,' and `shocked clump'
are labeled, and their spectra are shown in Figs.~\ref{fig:specrad}, 
\ref{fig:specac}, and~\ref{fig:manyspec},
respectively. The rectangle surrounding the `shocked clump' is the outline
of the region shown in Fig.~\ref{fig:CSmap}.
The radio continuum contours are linearly spaced at 5\%
intervals up to the maximum of the map (0.06 Jy/beam, Reynolds \& Moffett 1993).
\label{fig:comapamb}}

\figcaption[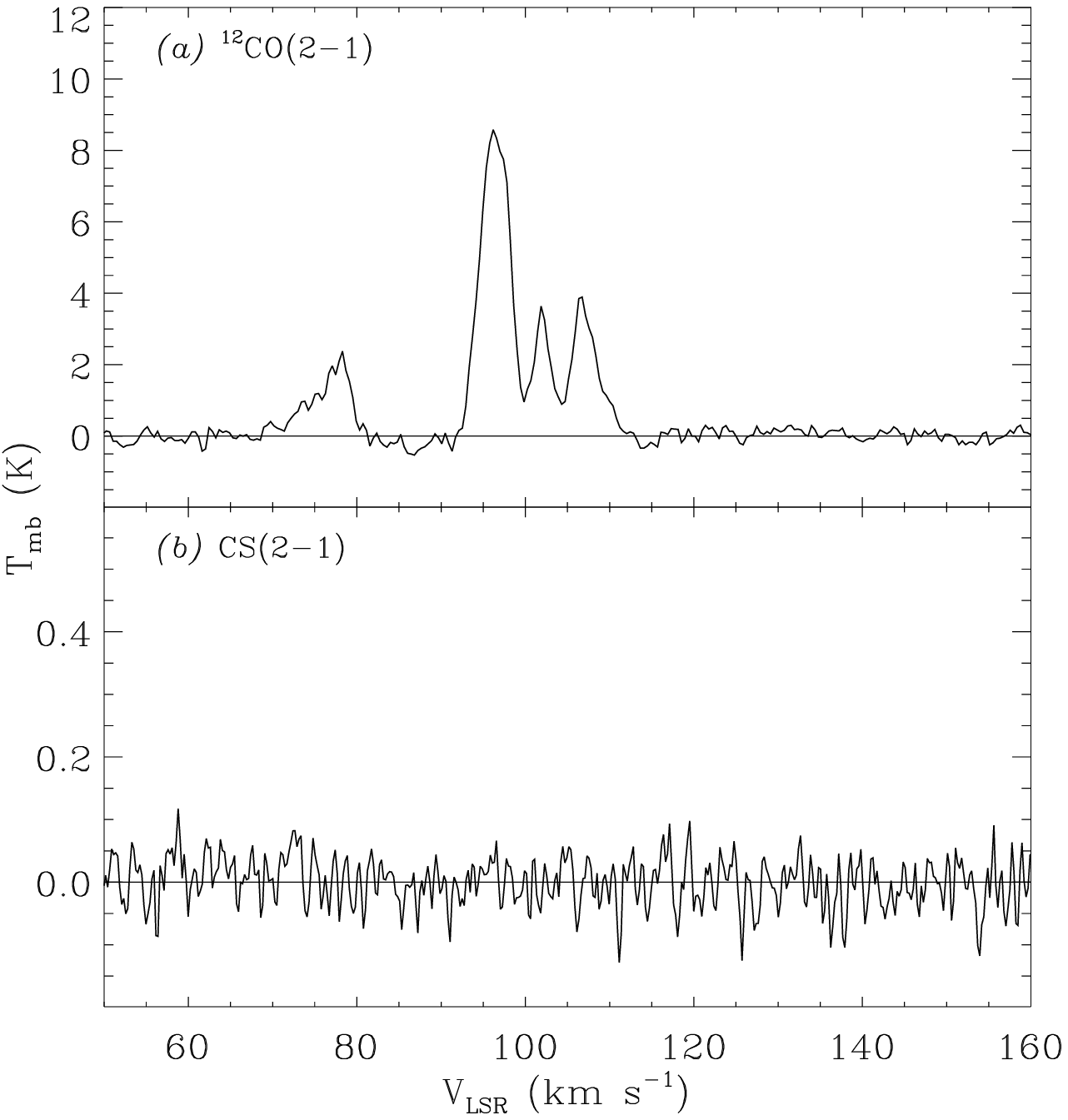]{Spectra of {\it (a)} $^{12}$CO($2\rightarrow 1$) and
{\it (b)} CS($2\rightarrow 1$) emission from the brightest part of the
radio shell of 3C~391. Each spectrum is the average of spectra within
$21^{\prime\prime}$ of relative position 
$(-130^{\prime\prime},+90^{\prime\prime})$.
\label{fig:specrad}}

\figcaption[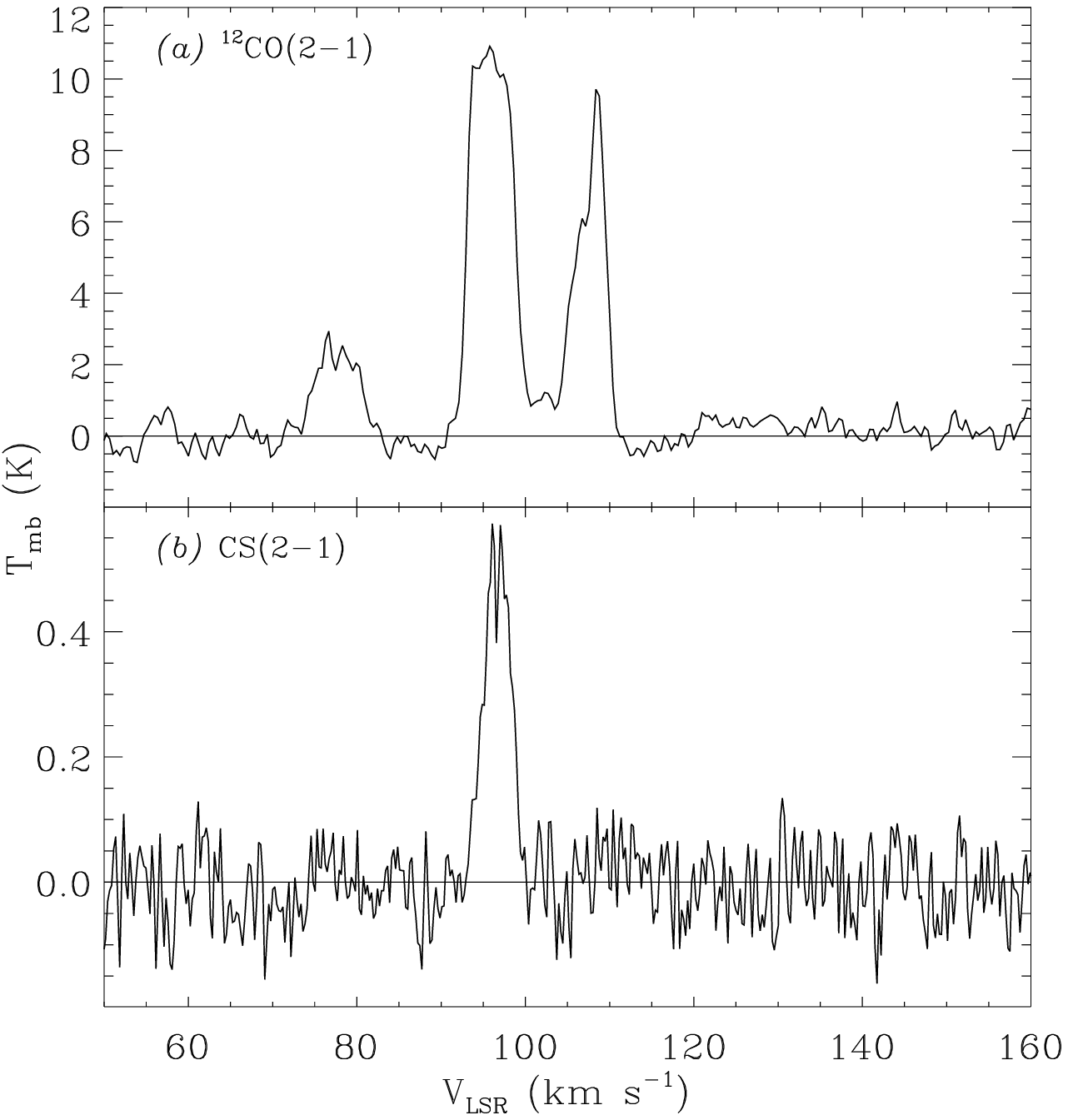]{Spectra of {\it (a)} $^{12}$CO($2\rightarrow 1$) and
{\it (b)} CS($2\rightarrow 1$) emission a relatively bright clump in the
ambient gas in the parent molecular cloud.
Each spectrum is the average of spectra within
$21^{\prime\prime}$ of relative position $(-150^{\prime\prime},+220^{\prime\prime})$.
\label{fig:specac}}

\figcaption[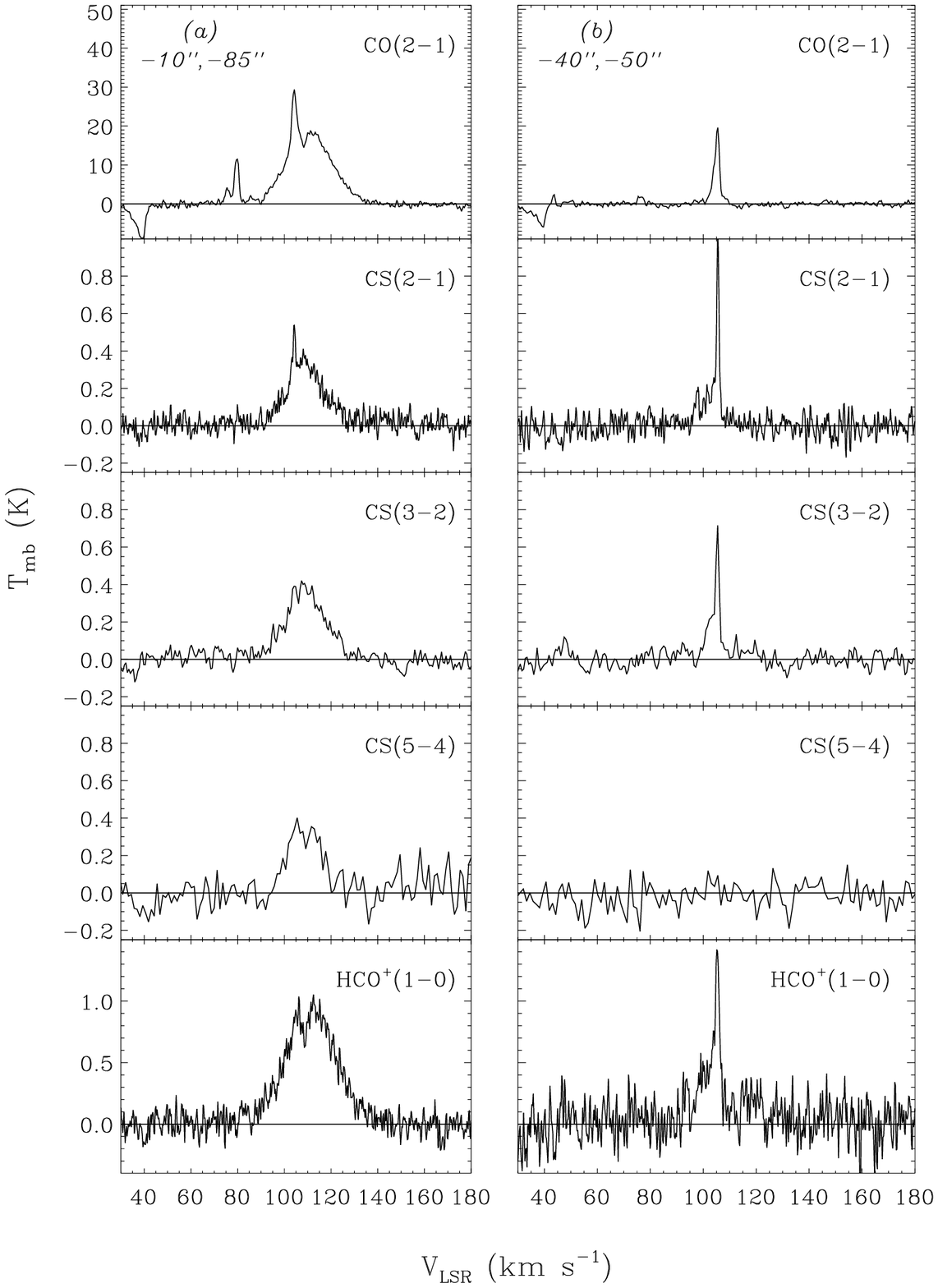]{
Spectra of the 3C~391 shocked molecular clump. 
The spectra in the right-hand panel {\it (a)}
are toward the peak of the wide-line emission of this clump,
at relative position $(-40^{\prime\prime},-50^{\prime\prime})$. 
The spectra on the right-hand panel {\it (b)}
are toward the peak of narrow-line emission from this clump, at
relative position $(-10^{\prime\prime},-85^{\prime\prime})$. 
Note that the wide emission lines are much brighter in the
high-excitation \csfive\ line, compared with the narrow-line region.
The spectra of the narrow-line region are similar to those of the
parent molecular cloud in width and relative line brightnesses.
\label{fig:manyspec}}

\figcaption[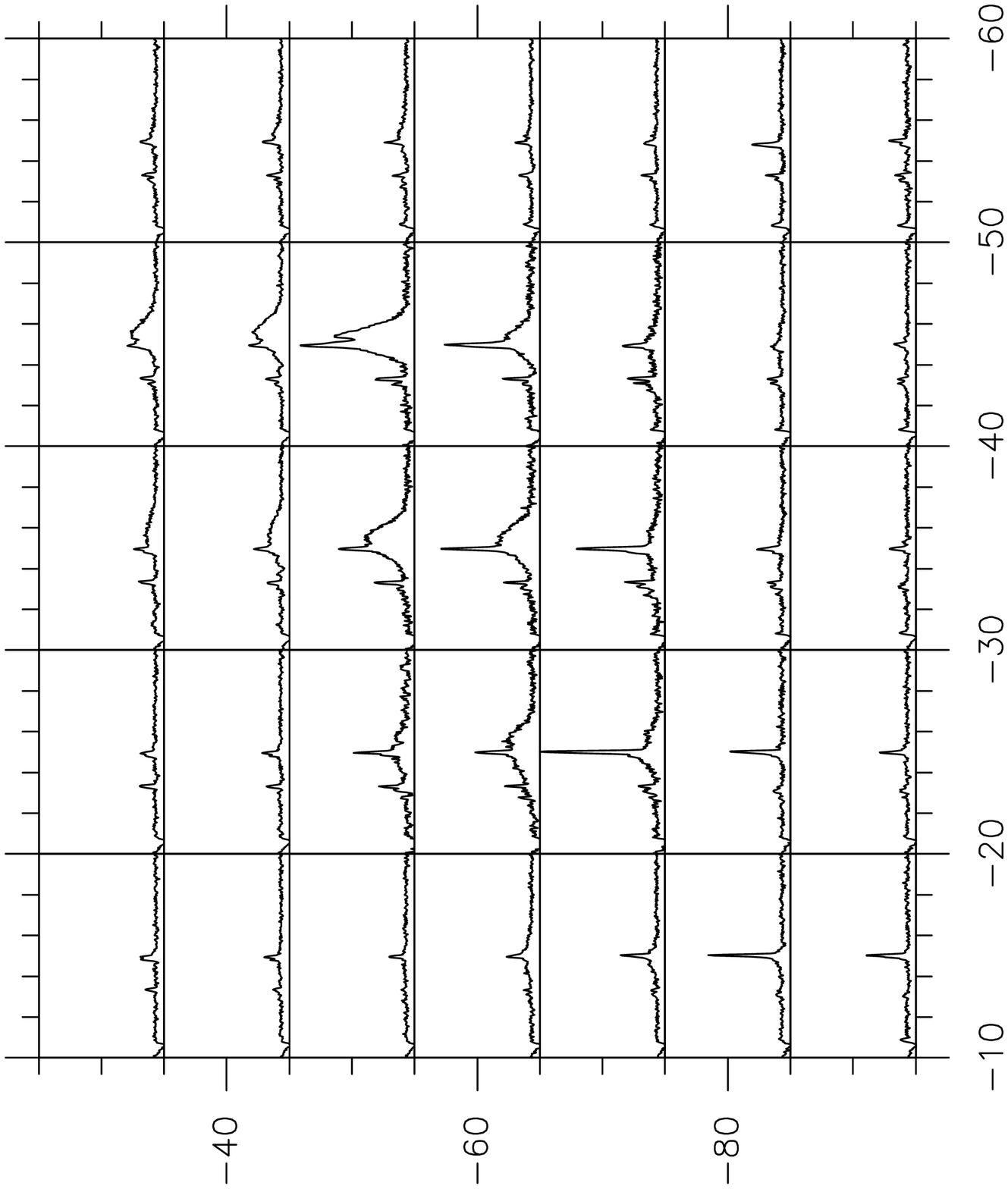]{Grid of $^{12}$CO($2\rightarrow 1$) spectra near Maser 1.
The spacing between spectra is $10^{\prime\prime}$, the velocity scale in each
box is from 30 to 180 \kms, and the main beam brightness temperature in each
box ranges from -4 to 62 K. The line profile changes rapidly from position
to position. The narrow component of the line profile is the pre-shock gas,
while the broad component (with a width of 20 \kms) is the shocked gas.
\label{fig:comosaic}}

\figcaption[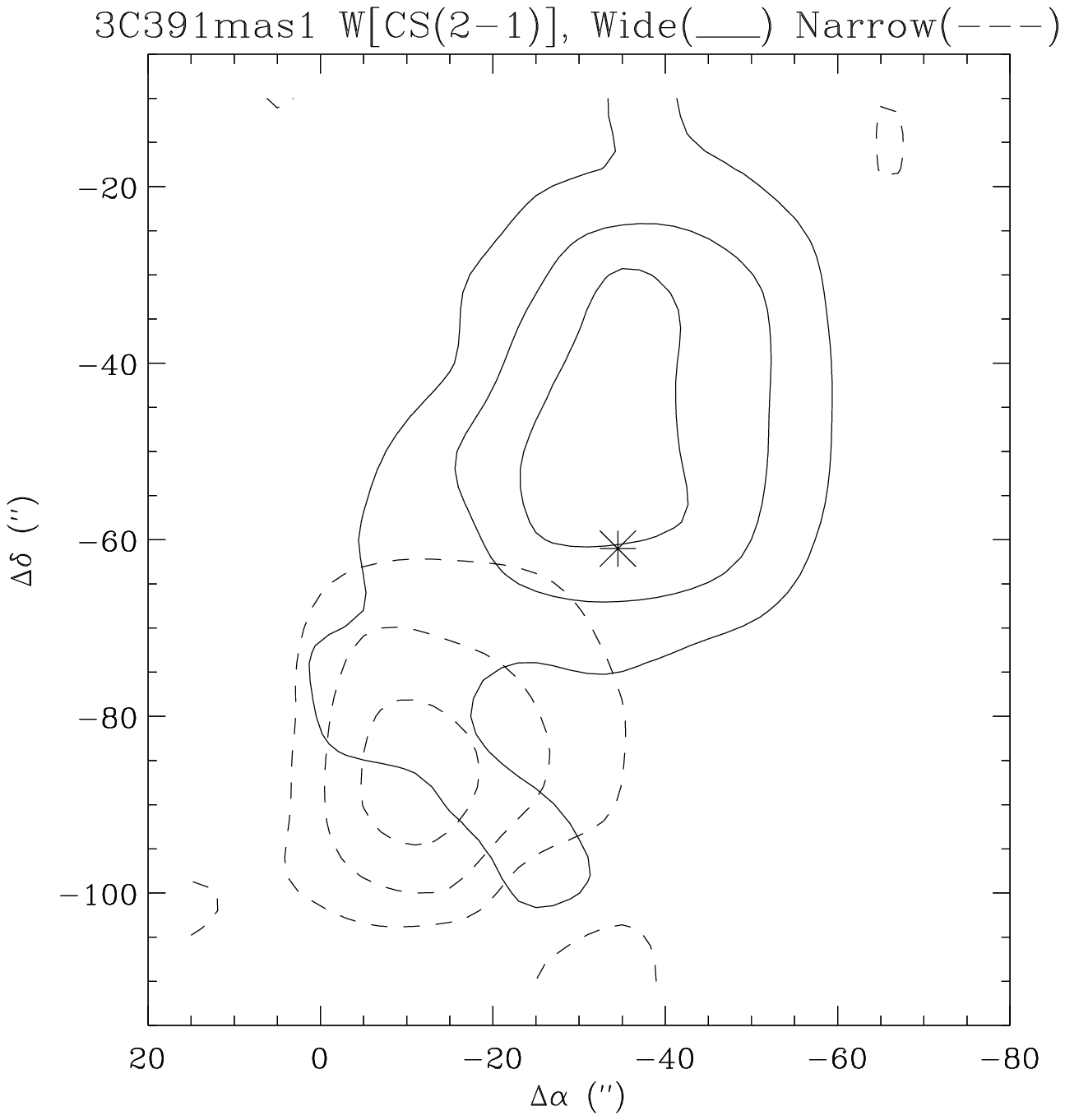]{A small map showing separately the brightnesses of the wide (solid contours)
and narrow (dashed contours) components of the CS($2\rightarrow 1$) lines
for the Maser 1 clump. The location of this region is illustrated in
Fig. 1.
Solid contours are drawn at 5.5, 11, and 16.5 K~km~s$^{-1}$,
and dashed contours are drawn at 1, 2, and 3 K~km~s$^{-1}$.
The position of the OH maser is indicated by an asterisk.\label{fig:CSmap}}

\figcaption[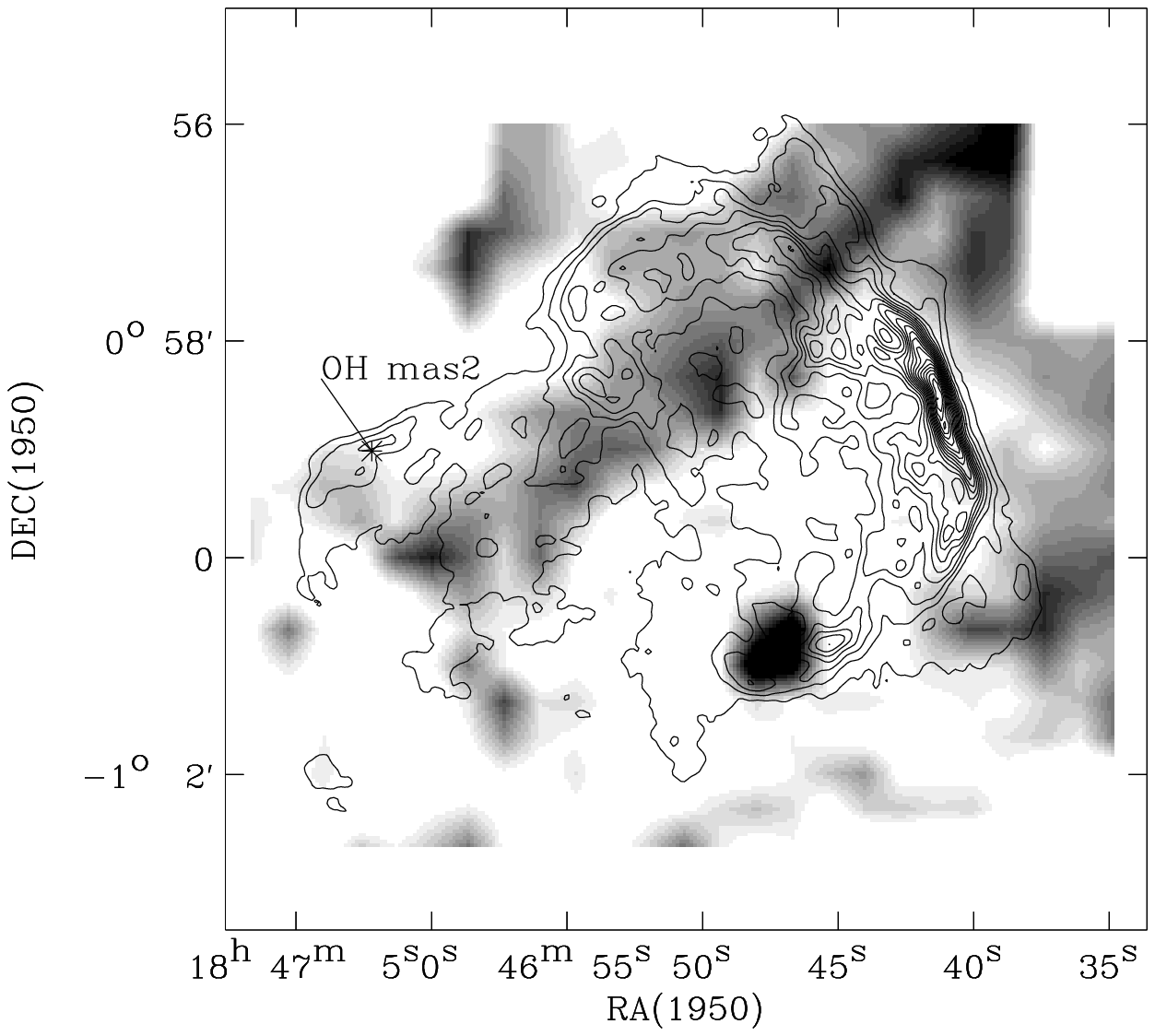]{Map of the \cotwo\ line integrated over the velocity range
108 to 110 \kms, overlaid on radio continuum contours as in Fig. 1. 
The map is dominated by emission from the parent cloud, in the 
northwest, and a long filament that runs from the parent cloud along
the northeastern edge of the radio shell. The location of the second OH 1720 MHz 
maser, whose velocity matches this gas, is labeled. \label{fig:comapshock}}

\figcaption[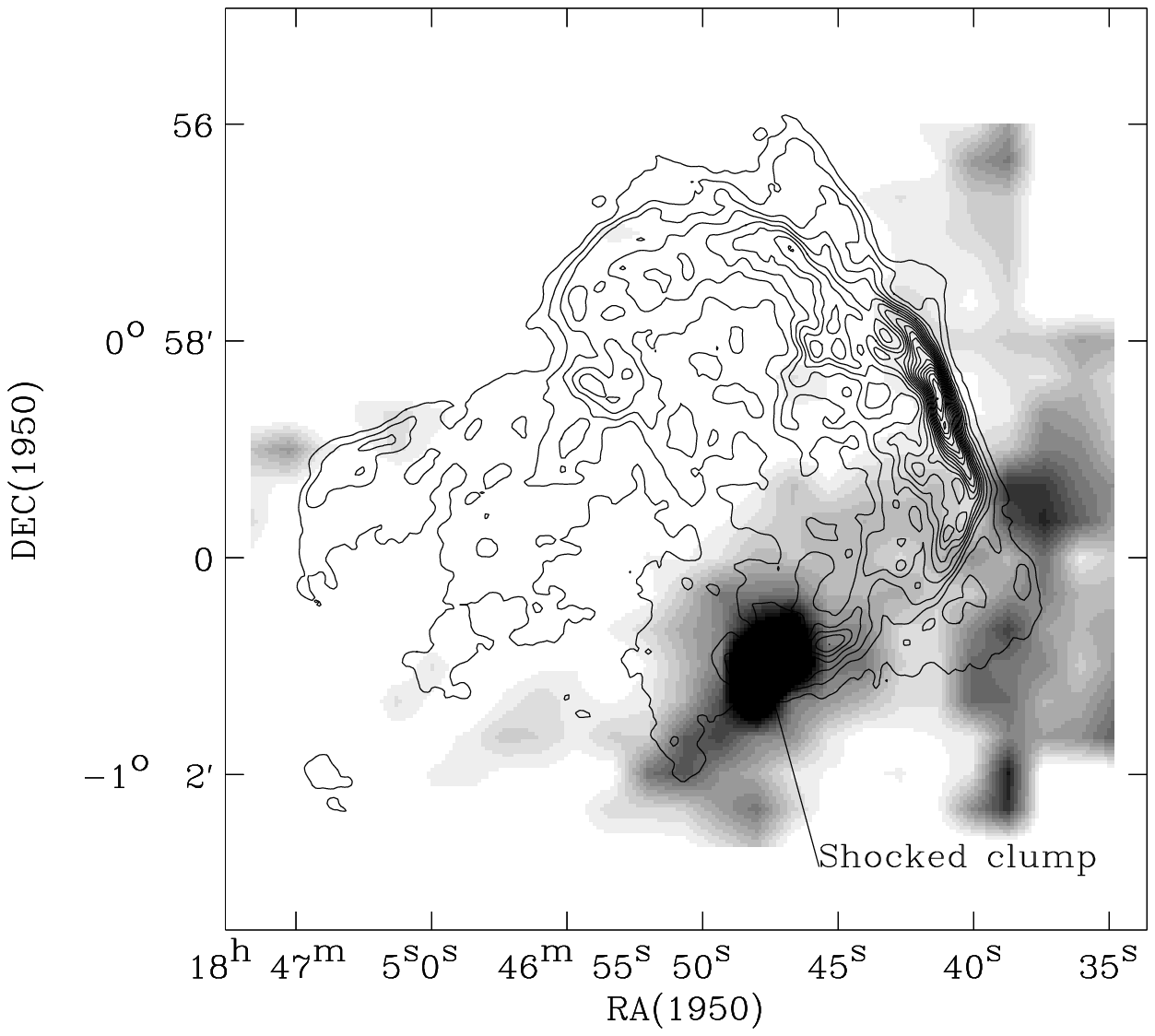]{Map of the \cotwo\ line integrated over the velocity range
104 to 106 \kms, overlaid on radio continuum contours as in Fig. 1.
The map is dominated by the shocked clump and some relatively
diffuse emission that lies roughly along the southwestern edge of the radio shell.
The brightest OH 1720 MHz maser in this remnant is located in the shocked clump,
and the OH and CO peak velocities agree.
\label{fig:comapshockb}}

\def\extra{
\figcaption[posvel.ps]{Velocity-position map of the CO($2\rightarrow 1$) emission from 3C~391,
along a cut starting (bottom) in the ambient molecular cloud, running through the
middle of the remnant, and emerging at the edge of the mapped region.}\label{fig:posvel}
}

\figcaption[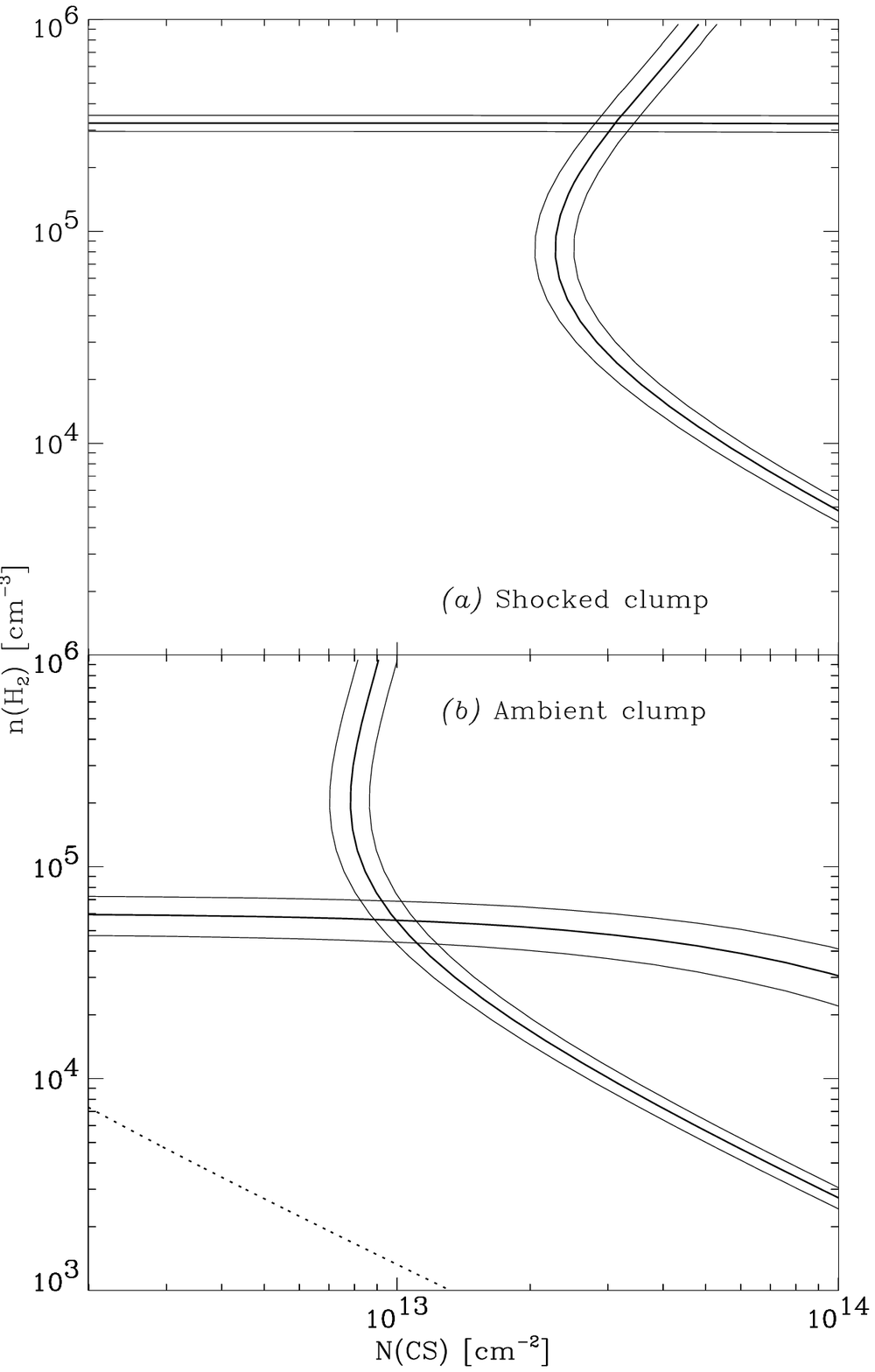]{
Excitation diagram for CS, showing the values of H$_2$ volume density and
CS column density of LVG calculations that satisfy the observed constraints.
Panel {\it (a)} shows the results for the shocked clump. The nearly horizontal
constraint shows models that predict the observed ratio of 
$5\rightarrow 4$ to $2\rightarrow 1$ line brightness, and the more
vertical constraint shows models that predict the observed brightness of the
$2\rightarrow 1$ line (assuming emission fills the beam).
Panel {\it (b)} shows the results for the ambient clump in the parent molecular
cloud. The nearly horizontal constraint shows models that predict the observed 
ratio of $3\rightarrow 2$ to $2\rightarrow 1$ line brightness, and the more
vertical constraint shows models that predict the observed brightness of the
$2\rightarrow 1$ line (assuming emission fills the beam).
Each curve is surrounded by similar curves, showing the effects of $\pm 10$\%
uncertainties in the observed quantitites.
The dotted line in panel {\it (b)} shows models that predict
$2\rightarrow 1$ line brightness equal to our detection limit, which
applies, for example, to the radio continuum ridge.
\label{fig:csexcite}}



\plotone{f1.ps}

\plotone{f2.eps}

\plotone{f3.eps}

\plotone{f4.ps}

\plotone{f5.eps}

\plotone{f6.ps}

\plotone{f7.ps}

\plotone{f8.ps}

\def\extra{
\plotone{posvel.ps}
}

\plotone{f9.eps}

\end{document}